\documentclass[a4paper]{article}

\usepackage{INTERSPEECH2021}
\usepackage{lipsum}
\usepackage{caption}
\usepackage[colorlinks=true,allcolors=blue]{hyperref}
\usepackage[noabbrev]{cleveref}
\usepackage{xcolor}
\usepackage{nicefrac}
\usepackage{multirow}
\usepackage[linesnumbered,ruled,vlined]{algorithm2e}
\usepackage{tikz}

\title{Federated Domain Adaptation for ASR with Full Self-Supervision}
\name{Junteng Jia, Jay Mahadeokar, Weiyi Zheng, Yuan Shangguan, Ozlem Kalinli, Frank Seide}
\address{Meta AI}
\email{\{juntengjia,jaym,wyz,yuansg,okalinli,seide\}@fb.com}

\newcommand*\circled[1]{\tikz[baseline=(char.base)]{\node[shape=circle,draw,inner sep=0.2pt] (char) {#1};}}
\definecolor{darkgreen}{RGB}{0,128,0}
\SetCommentSty{mycommfont}
\SetKwInput{KwInput}{Input}                % Set the Input
\SetKwInput{KwOutput}{Output}              % set the Output
\SetKwFor{Loop}{loop}{}{end}

\crefname{algocf}{algorithm}{algorithms}
\Crefname{algocf}{Algorithm}{Algorithms}

\begin{document}

\maketitle
% 

% \input{paper/abstract_ServerEmulatedFL.tex}
% \input{paper/introduction_ServerEmulatedFL.tex}

% JJ: trying to experience with another writing here
\begin{abstract}
Cross-device federated learning (FL) protects user privacy by collaboratively training a model on user devices, therefore eliminating the need for collecting, storing, and manually labeling user data. 
While important topics such as the FL training algorithm, non-IID-ness, and Differential Privacy have been well studied in the literature, this paper focuses on two challenges of practical importance for improving on-device ASR: the lack of ground-truth transcriptions and the scarcity of compute resource and network bandwidth on edge devices.
First, we propose a FL system for on-device ASR domain adaptation with full self-supervision, which uses self-labeling together with data augmentation and filtering techniques.
The system can improve a strong Emformer-Transducer based ASR model pretrained on out-of-domain data, using in-domain audio without any ground-truth transcriptions.
Second, to reduce the training cost, we propose a self-restricted RNN Transducer (SR-RNN-T) loss, a variant of alignment-restricted RNN-T that uses Viterbi alignments from self-supervision.
To further reduce the compute and network cost, we systematically explore adapting only a subset of weights in the Emformer-Transducer. 
Our best training recipe achieves a $12.9\%$ relative WER reduction over the strong out-of-domain baseline, which equals $70\%$ of the reduction achievable with full human supervision and centralized training.
\end{abstract}
\noindent\textbf{Index Terms}: ASR, Adaptation, Federated Learning
\footnote{This paper was submitted to Interspeech 2022.}
\section{Introduction}
With the increasing adoption of deep learning in artificial intelligence applications, data privacy is of growing concern~\cite{mireshghallah2020privacy,ha2020security}.
Automatic speech recognition (ASR) is a particularly sensitive use case due to the personalized nature of speech~\cite{schuller2013computational,cherkassky2021voice}.
Aiming to protect user privacy, cross-device federated learning (FL) has been proposed and applied to a variety of tasks~\cite{yang2018applied,hard2018federated,leroy2019federated,ramaswamy2019federated}.
FL is a distributed training paradigm that allows a loose federation of trainers to collectively improve a shared model~\cite{konevcny2016federated,mcmahan2017communication,nandury2021cross,li2021survey}.
%
%In cross-device FL, these trainers run on individual devices, thereby protecting user privacy since no data ever leaves the devices.
%

Cross-device FL for ASR faces several unique challenges.
These include the lack of ground truth transcriptions; the high costs of computation and network communication; the non independent and identical distribution of training data (non-IID-ness); and the difficulty of providing privacy guarantees. 
Several recent works have considered cross-device FL for ASR applications~\cite{dimitriadis2020federated,gao2021end,yu2021federated,guliani2021training,cui2021federated,paulik2021federated}.
%
% [Mention one line on FedAvgM and other algos that tackle distributed updates for ASR].
%
In particular, weighted model averaging~\cite{dimitriadis2020federated,gao2021end} and federated variation noise~\cite{guliani2021training} have been proposed to address the challenge of training on non-IID data, while differential privacy has been proposed to provide formal privacy guarantees~\cite{dwork2014algorithmic,abadi2016deep}.
In this paper, however, we address the other two practical challenges for on-device FL: 
the lack of ground truth transcriptions, and the scarcity of compute resources and network bandwidth.
%
% we will address other challenges in our future work.
%
%Combining with non-IID-ness and differential privacy will be explored in future work. 
% 
% We will address non-IIDness and differential privacy challengs in future work. 

\begin{figure}[t]
    \centering
    \includegraphics[width=1.0\linewidth]{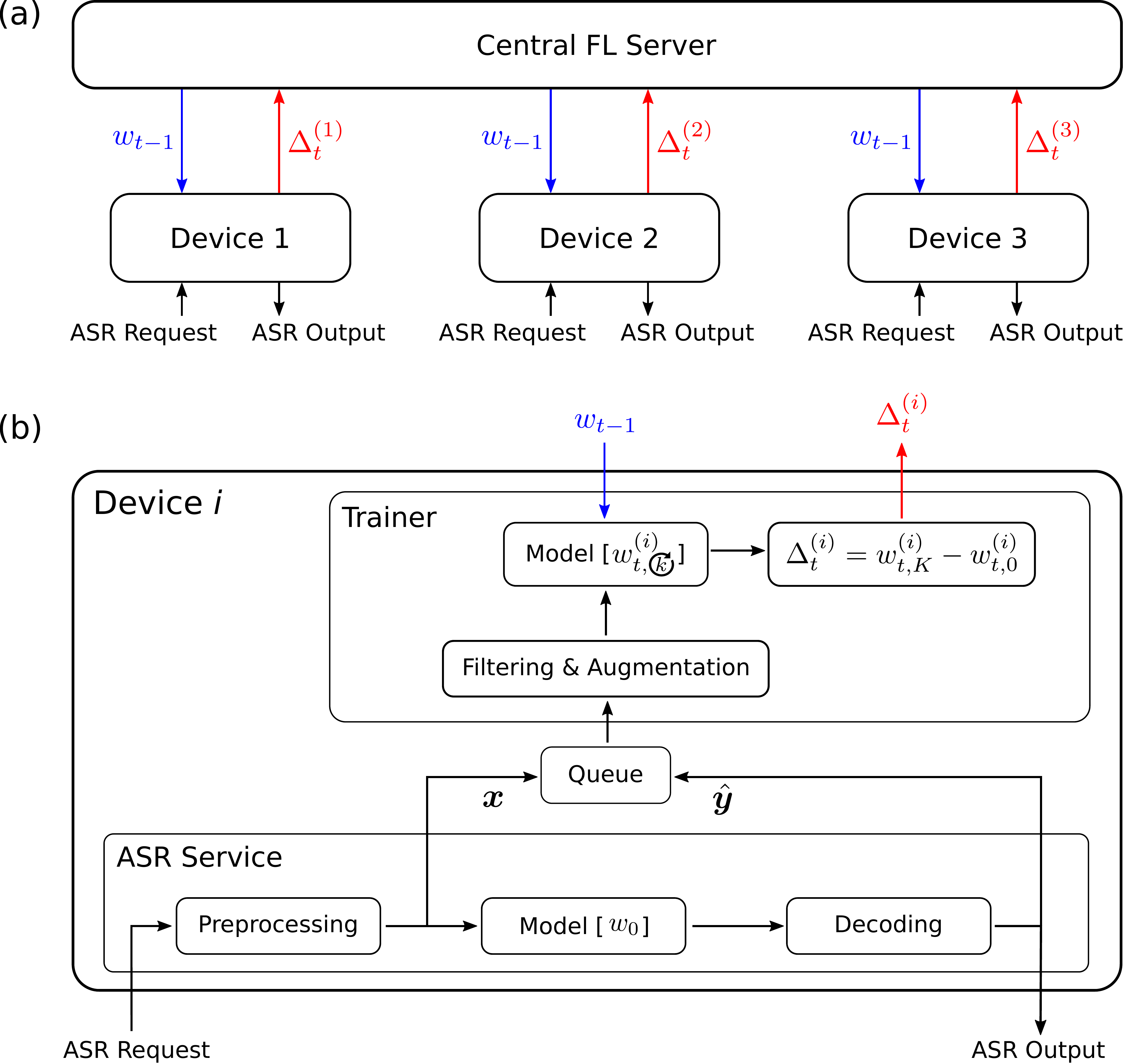}
    \caption{A FL system for on-device ASR domain adaptation with self-supervision. Each device runs a local trainer alongside the ASR service.}
    \label{fig:overview}
\end{figure}

A common assumption in existing literature on FL for ASR is the availability of ground-truth transcriptions~\cite{dimitriadis2020federated,gao2021end,yu2021federated,guliani2021training,cui2021federated,paulik2021federated}. 
However, this assumption is not practical. 
Users of on-device ASR have neither the incentive to manually transcribe their audio, nor the desire to frequently edit inaccurate automatic transcriptions. 
Assuming no labeled audio data, \cite{nandury2021cross} proposes to use a large pretrained teacher model to label target domain audio on cloud servers; other works, not focused on ASR, have studied semi-supervised FL for keyword spotting, audio classification~\cite{tsouvalas2021federated}, and image classification~\cite{jeong2020federated,he2021ssfl}.
However, to the best of our knowledge, there has been limited literature that explored on-device adaptation for ASR with self-supervision.
%
% However, to the best of our knowledge, literature on fully self-supervised on-device ASR domain adaptation is limited.
%
To this end, we propose an FL system for on-device model domain adaptation with self-supervision that runs alongside an ASR application.

On a high level, user audio and respective automatic transcriptions ({\em pseudo labels}) from an on-device ASR model are stored temporarily on the device in an encrypted queue.
Then, a trainer that runs alongside the ASR process consumes minibatches of examples from the queue and performs local model updates.
Model updates are periodically sent to a central FL server for aggregation.
We develop data augmentation and filtering techniques to improve robustness of the self-supervised learning.
Our offline experiments on two diverse FL adaptation scenarios confirm the effectiveness of the proposed system.

%
%and show that despite these challenges, our proposed FL system is able to improve a strong out-of-domain ASR.
%
% We show that despite this challenges, our proposed FL system is able to improve a strong out-of-domain ASR model on user on-device data without using any human supervision.
%
%
Another important challenge in cross-device FL is the cost of training and synchronization with the central FL server~\cite{mcmahan2017communication}, since edge devices typically have limited compute, battery, and network resources~\cite{konevcny2016federated}.
In this work, we present a combination of techniques to improve an Emformer-Transducer based ASR model~\cite{he2019streaming,rao2017exploring}, which is trained with the RNN-T loss~\cite{graves2012sequence}.
First, to reduce the compute cost during on-device adaptation, we propose a self-restricted (SR)-RNN-T loss, which extends the alignment-restricted (AR)-RNN-T~\cite{mahadeokar2021alignment} by replacing the external alignment with a Viterbi alignment (similar to \cite{kim2021reducing}) from the model itself.
Secondly, to further reduce the network cost during synchronization, we consider only adapting a subset of model weights.
Inspired by a recent work on speaker adaptation~\cite{huang2021rapid}, we study which subset of weights in the ASR model would give the best domain-adaptation performance.
We find that adapting the key/value matrices in the self-attention module can match (or surpass) full model adaptation, even though those matrices only account for $5.7\%$ model weights.
A related approach called partial variation training adapts different subset of weights on different devices~\cite{yang2021partial}.% however, our initial results find it less effective for our use case.

Combining all techniques, we achieve 12.9\% relative WER reduction for on-device adaptation --- 70\% of the reduction achievable with human supervision and centralized training.

\section{Methodology \label{sec:method}} 
\begin{algorithm}[t]
\DontPrintSemicolon
    \KwInput{pretrained model $w_{0}$, \# of rounds $T$, \# of devices $N$, \# of updates per round $K$}
    \KwOutput{finetuned model $w_{T}$}
    \vspace{0.03in}

    \SetKwProg{Fn}{Procedure}{:}{}
    \SetKwFunction{FCentralFLServer}{CentralFLServer}
    \SetKwFunction{FDevice}{Device}
    \SetKwFunction{FASRService}{ASRService}
    \SetKwFunction{FTrainer}{Trainer}
  
    \Fn{\FCentralFLServer{}}{
        \For{t $\in$ 1 \ldots T}{
            \For{i $\in$ 1 \ldots N}{
                $\textcolor{blue}{\texttt{SendModel}}(w_{t-1}, i, t)$\;
                $\Delta_{t}^{(i)} \gets \textcolor{red}{\texttt{ReceiveUpdate}}(i, t)$
            }
            $\Delta_{t} \gets \textstyle \frac{1}{N} \sum_{i=1}^{N} \Delta_{t}^{(i)}$\;
            $w_{t} \gets w_{t-1} + \beta (w_{t-1} - w_{t-2}) + \Delta_{t}$
        }
    }
    \vspace{0.03in}
  
    \Fn{\FDevice{$i$}}{
        $Q_{i} \gets \{\}$\;
        $\texttt{ASRService}(Q_{i})$\;
        $\texttt{Trainer}(i, Q_{i})$
    }
    \vspace{0.03in}
    
    \Fn{\FASRService{$Q_{i}$}}{
        \Loop{}{
            $\bm{x} \gets \texttt{GetRequest}()$\;
            $\hat{\bm{y}} \gets \texttt{EncodeDecode}(w_{0}, \bm{x})$\;
%           $\texttt{ReturnRequest}(\hat{y})$\;
            $\texttt{PushToQueue}(Q_{i}, (\bm{x}, \hat{\bm{y}}))$
        }        
    }
    \vspace{0.03in}
  
    \Fn{\FTrainer{$i$, $Q_{i}$}}{
        \For{t $\in$ 1 \ldots T}{
            $w_{t-1} \gets \textcolor{blue}{\texttt{ReceiveModel}}(i, t)$\;
            $w_{t,0}^{(i)} \gets w_{t-1}$\;
            \For{k $\in$ 1 \ldots K}{
                $B_{k} \gets \texttt{PopBatchFromQueue}(Q_{i})$\;
                $\bar{B}_{k} \gets \texttt{FilterAndAugment}(B_{k})$\;
                $w_{t,k}^{(i)} \gets \texttt{ModelUpdate}(w_{t,k-1}^{(i)}, \bar{B}_{k})$
            }
            $\Delta_{t}^{(i)} \gets w_{t,K}^{(i)} - w_{t,0}^{(i)}$\;
            $\textcolor{red}{\texttt{SendUpdate}}(\Delta_{t}^{(i)}, i, t)$
        }
    }
\caption{FL system for ASR domain adaptation.}
\label{alg:system}
\end{algorithm}
\subsection{Federated Learning Alongside the ASR Process \label{subsec:system}}
This section presents the design of the proposed FL system.
An initial ASR model is pretrained on large amounts of supervised data from a broad range of sources, using ground truth transcriptions or high-quality machine generated transcriptions from a large offline teacher model. 
Then, the FL system adapts the pretrained ASR model to target domain data using pseudo labels generated by the pretrained model itself, via a hierarchical SGD method known in the FL community as {\em Federated Averaging with Momentum}, abbreviated as FedAvgM~\cite{hsu2019measuring}, although the same method was first developed in the speech community under the name {\em blockwise model-update filtering}, or BMUF \cite{chen2016scalable}.

%
% In particular, we design the FL model updates to be part of the ASR on-device application. 
%
% When an ASR request are completed, we immediately pass the audio and the ASR output transcription, which are still held in RAM, to the FL trainer for performing a local gradient update.
% %
% The algorithm is the same as true on-device FL, except that, since no human transcription is needed and ASR workers are not limited to run during night hours while charging and connected to Wifi, learning can be performed on-the-fly right after ASR has completed, and the audio and transcript can be discarded immediately after completing the gradient computation.
% %
% On the other hand, since ASR servers process incoming utterances with random assignment, we do not need to be concerned with non-IID-ness of batches for now.

\Cref{fig:overview} illustrates the cross-device FL system, comprising a central FL server and many ASR devices, implementing \Cref{alg:system}.
Each device $i$ executes an ASR process that runs the pretrained model $w_{0}$, which is fixed throughout the FL process.
A device-specific trainer manages a second copy of the ASR model $w_{t,k}^{(i)}$ that is updated via FL.
Each time the device receives an ASR request, the pretrained model is used to transcribe audio $\bm{x}$ to generate transcription $\hat{\bm{y}}$, which we refer as \textit{pseudo labels} (this differs from~\cite{nandury2021cross} which uses a large offline teacher model for this purpose).
The $(\bm{x}, \hat{\bm{y}})$ pairs are added to a queue for later consumption by the trainer.
In each local model update, the trainer samples a minibatch from the queue and computes model gradients with the SR-RNN-T loss.
We find that it is important for adaptation performance to use confidence filtering and data augmentation.
Finally, after $K$ updates, the difference $\Delta^{(i)}$ is sent back to the central FL server for aggregation.

%
%This self-training setup would be necessary for on-device FL where the computation resource is limited.
%
%Next, we report some key designs that makes this setup effective.
%
% Later, we will discuss how we further reduce the computation and communication cost for on-device FL.

\subsection{Efficient RNN-T Adaptation with Self-Alignment \label{subsec:effectiveness}}
\begin{figure}[t]
    \centering
    \includegraphics[width=1.0\linewidth]{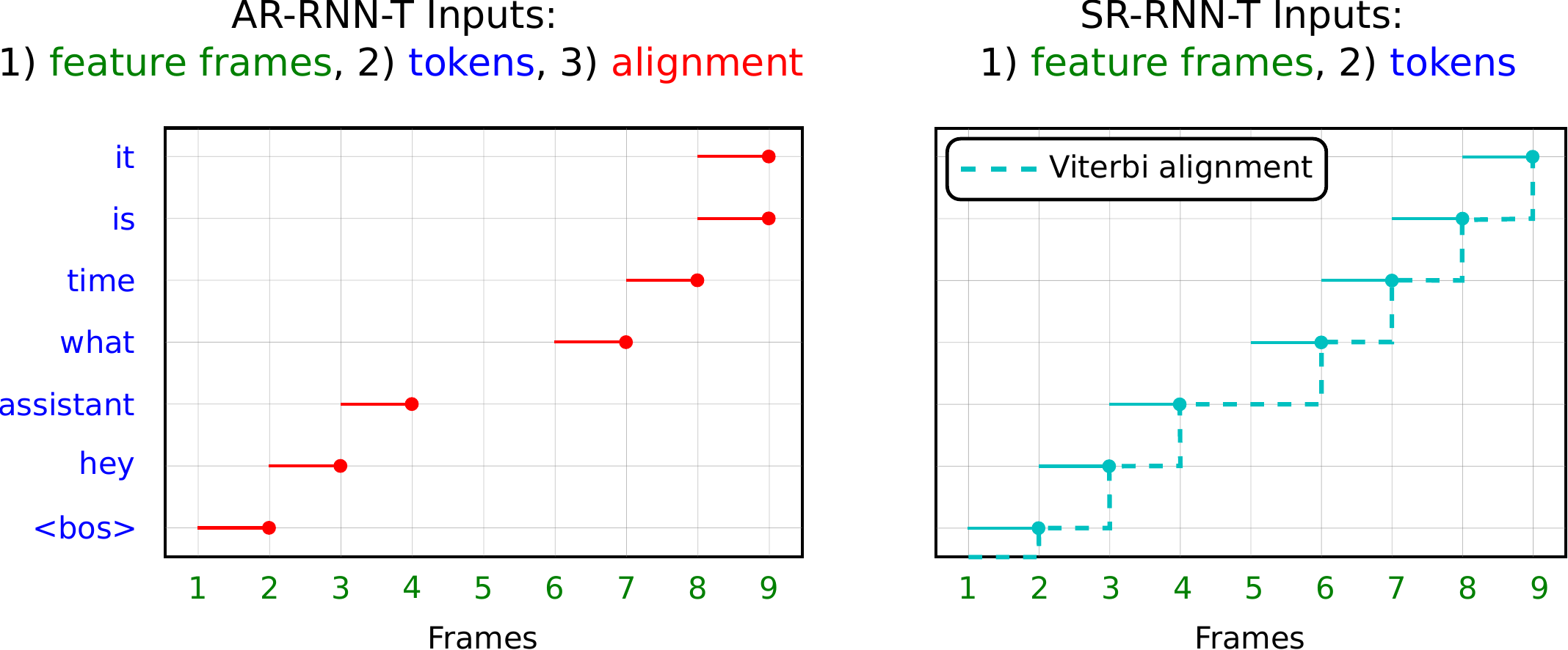}
    \caption{Illustration of the proposed SR-RNN-T loss. In contrast with AR-RNN-T, which relies on an external alignment (red), SR-RNN-T uses the Viterbi forced-alignment (cyan).}
    \label{fig:srrnnt}
\end{figure}
%
%In this section we describe the combination of techniques that we used to adapt the pretrained ASR model on new domains with federated learning. 
%
%One of the key realistic challenges for FL is the lack of supervised training data or transcriptions. We explore techniques to adapt the model using self-supervision.

%
We use an Emformer-Transducer~\cite{graves2012sequence,shi2021emformer} as our ASR model, pretrained with AR-RNN-T loss and centralized distributed data-parallel (DDP) training.
Let $\bm{x}$ denote the audio features over $T$ time steps and $\bm{y}$ denote the reference text over $U$ target tokens.
First, the encoder maps the audio features $\bm{x}$ into input embeddings $\bm{h}$, and the predictor maps the reference tokens $\bm{y}$ into output embeddings $\bm{g}$.
Then, a joiner network combines every pair of $h_{t}, g_{u} \in \bm{h}, \bm{g}$ to compute the probability distribution $\textrm{Pr}(\cdot|h_{t}, g_{u})$ over all possible output tokens including blank. 
During training, the RNN-T loss optimizes the sum of the probabilities over all possible audio-text alignments.
During inference, a beam search algorithm is used to output the most probable sequence, which we refer to as $\hat{\bm{y}}$.

In FL scenario, $\bm{y}$ is not available.
Instead, we use the most probable sequence $\hat{\bm{y}}$ from inference as target labels.
A straight-forward implementation of the RNN-T loss would directly use $\bm{x}$ and $\hat{\bm{y}}$ and optimize over all possible alignments.
However, this approach is computationally intensive.
As a potential alternative, \cite{mahadeokar2021alignment} showed that the AR-RNN-T loss can greatly reduce memory and compute by restricting the forward-backward algorithm to a band around a ground truth alignment.
However, AR-RNN-T relies on external ground-truth alignments generated offline with a separate hybrid model.
Motivated by some recent studies~\cite{kim2021reducing}, we propose to use Viterbi forced-alignment from the model itself in place of an external alignment.
Particularly, the forced-alignments are computed on the fly over the RNN-T grid using the backtracing (see Fig.~\ref{fig:srrnnt}).
We refer the resulting loss function as self-restricted (SR)-RNN-T loss.
In our design of the FL system, the Viterbi forced-alignment is further approximated with the most probable alignment encountered during beam search decoding after pruning.

\subsection{Data Filtering and Augmentation \label{subsec:filtering_and_augmentation}}
One main concern with self-supervision is the quality of the pseudo labels $\hat{\bm{y}}$.
Incorrect pseudo labels carry the risk of reinforcing errors.
We find an effective mitigation to be data filtering, where
all utterances with $\log{\textrm{Pr}(\hat{\bm{y}})} < \theta$ are discarded.
The threshold $\theta$ is a tunable hyper-parameter, which is empirically determined on a small development dataset.
We find that data filtering is especially important if adaptation domain data is noisy (see \cref{subsec:video}).
%
% For instance, in the video adaptation experiment, the pretrained model has a high WER on the noisy target domain videos, therefore filtering is critical (see~\cref{subsec:video}).

On the other hand, when the target domain audios are clean, we find the semi-supervised self-training can be further boosted with data augmentation (speed perturbation and additive noise) while keeping the pseudo labels unchanged (see \cref{subsec:device}).

\subsection{Finetuning Only a Subset of Model Weights \label{subsec:efficiency}}
%
%Since FL requires periodic synchronization of the model updates, communication efficiency is another critical problem, especially for the cross-device FL setting where some devices might not have fast internet connection.
%
To reduce the cost of communication (and computation), we consider only adapting and communicating a subset of the model weights.
Inspired by a recent study~\cite{huang2021rapid}, we systemically studied which subset of weights in the Emformer-Transducer model provides the best trade-off between domain-adaptation performance vs.~communication cost. 
We explore adaptation of encoder, predictor and joiner separately.
Within the encoder, we carefully analyze adapting only attention or attention key/value matrices.  
We find adapting the key/value matrices is very effective for the target domain, and freezing the rest of weights serves as a regularization technique, which prevents catastrophic forgetting on the source domain.
%
% We were pleasantly surprised by a small accuracy improvement in our of our tasks.
%
% Interestingly, only adapting the key/value matrices in the self-attention module of the encoder can achieve (or surpass) full model adaptation performance on the target domain, even though those matrices only account for $5.7\%$ of model weights in our experiments.
%
%Furthermore, we find that freezing the rest of weights also serves as an effective regularization technique, which prevents catastrophic forgetting on the source domain.

% \begin{itemize}
%     \item sr-rnnt loss function
%     \begin{itemize}
%         \item we use self-alignment restricted rnnt (sr-rnnt) to reduce cost
%     \end{itemize}
%     \item finetuning on a subset of parameters
%     \begin{itemize}
%         \item reduce communication cost
%         \item prevent over-fitting on a small finetuning dataset
%         \item best is to finetune attention-kv or bias+layernorm
%     \end{itemize}
% \end{itemize}

% \subsection{Others}
% From discussion with frank, talk about other methods that we don't discuss. 
% Need to show data parallelization vs BMUF comparison
\section{Experiments}\label{sec:experiments}
% Naming of dataset: 
% Stella-Ph2 is Vendor Collected
%
\vspace{-0.15in}
\begin{table}[ht!]
    \centering
    \caption{Different experiment setups for simulating diverse set of FL experiments. The 1st row in each setup denotes the source domain data used for model pretraining, and 2nd row denotes the target domain data used for FL adaptation.}
    \resizebox{1.0\linewidth}{!}{
    \begin{tabular}{r|rrr}
    \toprule
    \textbf{Setup}              & \textbf{Dataset} & \textbf{Train (hours)} & \textbf{Eval (utterances)} \\
    \midrule
    \multirow{2}{*}{Device}     & Mixture          & 1.5M                 & 3.6K  \\
                                & Edge             & 9.7K                 & 28.6K \\
    \midrule
    \multirow{2}{*}{Video}      & Clean            & 17.4K                & 1.2K \\
                                & Noisy            & 1.2K                 & 1.5K \\
%   \midrule
%   \multirow{2}{*}{Public}     & LibriSpeech      & 5.8K  \\
%                               & WSJ              &  73   \\
    \bottomrule
    \end{tabular}
    }
    \label{tab:datasets}
\end{table}
\vspace{-0.20in}
\subsection{Data Setup}
To simulate realistic federated learning in practical scenarios, we consider two different experiment setups.
Each setup takes a converged model pretrained on a source domain and performs FL adaptation to a target domain.
The adapted model is tested on the evaluation data from both the source and target domains.
\Cref{tab:datasets} gives more information on the data used in each setup. \\[-0.1in]

\noindent \textbf{Device:}
% \textbf{Device setup:}
%
This setup closely resembles a cross-device FL use case.
For pretraining, we use a mixture of data sources, including public videos, audios collected by paid third-party vendors on portal, and synthetic audios generated by text-to-speech.
This dataset contains 1.5 Million hours of audios, which are transcribed either through human annotators or a large teacher model (see~\cite{xiao2021scaling} for details).
%
% Both types of transcriptions are considered as the ground truth.
%
For adaptation, we use assistant and dictation audios recorded on edge devices, which are collected by external vendors under clean acoustic conditions.
The source domain evaluation data is a separately prepared conversation dataset, while the target domain evaluation data is a held out subset of the on-device recordings. \\[-0.1in]

\noindent \textbf{Video:} 
%
%This setup is different from the device adaptation, since the target domain contains more diverse noisy audios. 
%
For pretraining, we use 17.4K hours of long form videos (averaged 45 sec) with relative clean audios.
For adaptation, we use short form videos (averaged 12 sec) that might be recorded in noisy environments.
%
% modified per legal request from SRT
Both datasets are de-identified before transcription by human annotators.
%
% Both datasets are transcribed by human annotators.
%
A subset of the audios from both the clean and noisy videos are held out for evaluation.

% \noindent \textbf{Public:}
% %
% In this setup, we use the publicly available LibriSpeech~\cite{panayotov2015librispeech} dataset for pretraining and Wall-street-journal (WSJ) dataset for adaptation.
% %
% On Librispeech, we further apply noise augmentation and speed distortion, similar to~\cite{mahadeokar2021alignment}, to create 5.8K hours of training dataset from the 960 hours of available read English speech.

\subsection{Modeling}
Our ASR model is an Emformer-Transducer~\cite{shi2021emformer}.
Features are 80 mel-spaced log filter-bank channels with a sliding window of 25ms at 10ms frame shift, regularized with SpecAugment~\cite{park2019specaugment}.
A time reduction layer stacks features with a stride of 6, and the resulting 480-dimensional vectors form the input to the encoder.
The predictor is a 2-layer LSTM model with a hidden dimension of 256.
Both the encoder and the predictor outputs are projected to 768-dimension tensors and combined in the RNN-T joiner, which consists of a ReLU followed by a softmax layer with a sentence-piece vocabulary size of 4096.

For each task, we first pretrain the ASR model on the source domain until convergence, using an AR-RNN-T loss with buffer size $b_{l}=5, b_{r}=25$ and an Adam optimizer with learning rate $10^{-3}$ and tri-stage learning rate scheduling.
We then finetune the ASR model to the target domain using 32 parallel workers.
Each worker uses an Adam optimizer with a fixed learning rate of $10^{-4}$ for local model updates.
The local models on different workers are synchronized every 20 local updates using FedAvgM/BMUF with a block momentum of 0.8.

\subsection{Device Adaptation \label{subsec:device}}
\begin{figure}[t]
    \centering
    \includegraphics[width=1.0\linewidth]{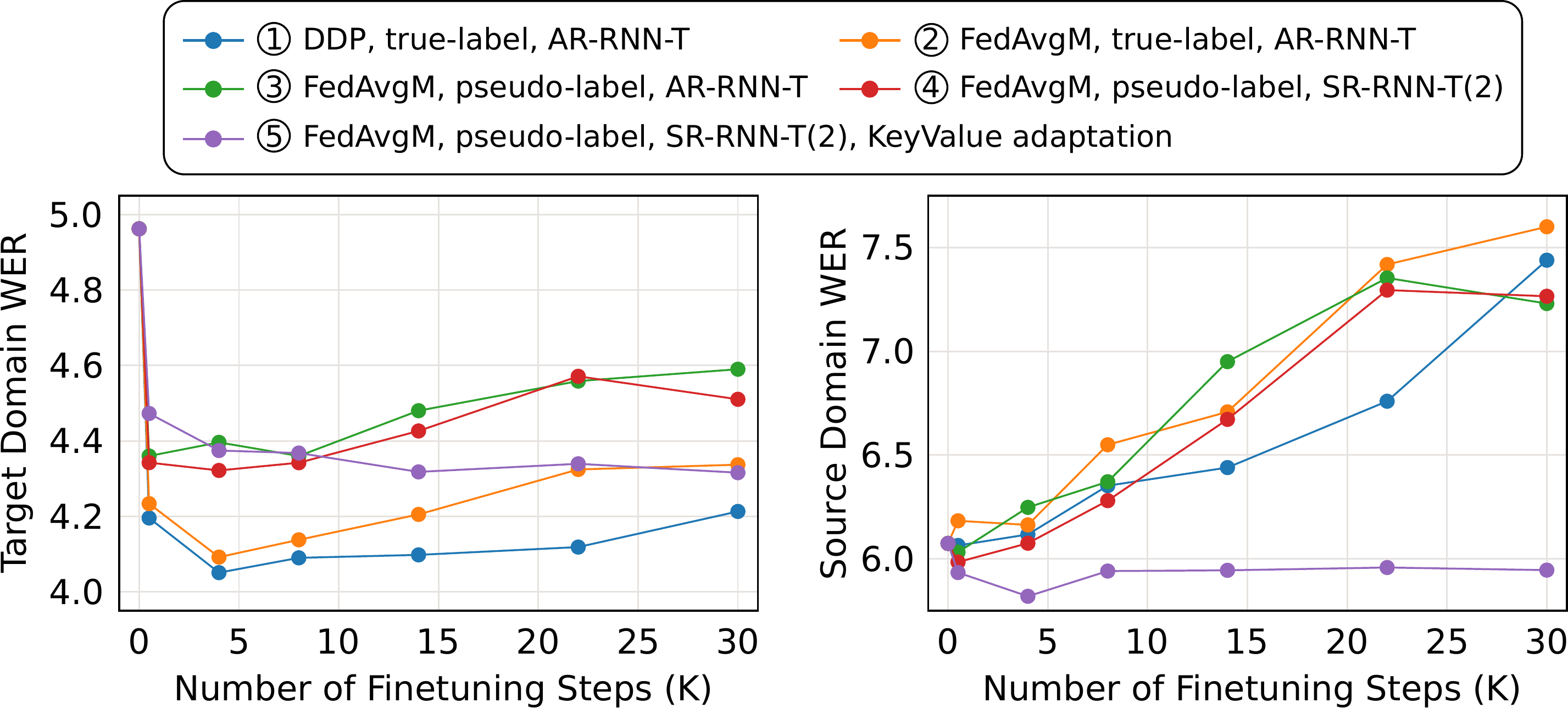}
    \caption{
    Convergence of the Device adaptation experiments. The combination of FedAvgM/BMUF, pseudo-labels, SR-RNN-T and KeyValue adaptation not only achieves a low WER on the target domain, but also prevents catastrophic forgetting on the source domain.
    }
    \label{fig:device}
\end{figure}
The Device adaptation setup adapts from the source domain ``Mixture'' to the target domain ``Edge''.
The baseline pretrained model achieves a lower WER on the target domain (4.96) than on the source domain (6.07), since the latter captures diverse acoustic conditions while the former is a commissioned data collection that is mostly clean. \\[-0.1in]
\begin{table}[h!]
  \caption{Summary of Device adaptation experiments.}
  \label{tab:device}
  \centering
  \begin{tabular}{l|rr}
    \toprule
    \textbf{Model}                                      & \textbf{Target} & \textbf{Source} \\
    \midrule
    Pretrained                                          & 4.96            & 6.07 \\
    \midrule
    $\circled{1}$ (DDP, True Labels, AR-RNN-T)          & 4.05            & 6.12 \\
    $\circled{2}$ ($\circled{1}$ + FedAvgM/BMUF)        & 4.09            & 6.16 \\
    $\circled{3}$ ($\circled{2}$ + Pseudo Labels)       & 4.35            & 6.17 \\
    $\circled{4}$ ($\circled{3}$ + SR-RNN-T)            & 4.32            & 6.08 \\
    $\circled{5}$ ($\circled{4}$ + KeyValue Adaptation) & 4.32            & 5.95 \\
    \bottomrule
  \end{tabular}
\end{table}

\noindent \textbf{Lower bound:}
We first consider supervised adaptation with centralized DDP training, using the ground-truth labels and the AR-RNN-T loss.
The adapted model's WER on the source and target domains across different checkpoints are plotted in \cref{fig:device}~$\circled{1}$.
In particular, the best checkpoint (see \cref{tab:device}~$\circled{1}$) reduces the WER $4.96 \rightarrow 4.05$ on the target domain, which serves as a lower bound for semi-supervised FL adaptations.
%
% This can be addressed by mixing server side updates using seed-model training data (results not shown in this paper).
%
We also observe a slight WER increase $6.07 \rightarrow 6.12$ on the source domain. \\[-0.1in]

\noindent \textbf{FedAvgM/BMUF:}
Next, we look at the effect of using the FL training algorithm, by replacing DDP with FedAvgM/BMUF while keeping the labels and the loss function the same ($\circled{2}$).
We observe a marginal WER increase on both the target domain (to $4.09$) and the source domain (to $6.16$).
\begin{table}[h]
  \caption{The effect of data augmentation. When target domain is clean, adaptation is less effective without data augmentation.}
  \label{tab:augmentation}
  \centering
  \begin{tabular}{l|rr}
    \toprule
    \textbf{Model}                     & \textbf{Target} & \textbf{Source} \\
    \midrule
    $\circled{3}$                      & 4.35                  & 6.17 \\
    $\circled{3}$ - Data Augment       & 4.48                  & 6.44 \\
    $\circled{3}$ - Data/Spec Augment  & 4.89                  & 6.67 \\
    \bottomrule
  \end{tabular}
\end{table}

\noindent \textbf{Pseudo labels:}
Now, we introduce pseudo labels to determine the effects of self-supervision ($\circled{3}$).
Although not as effective as the true labels, self-supervision still reduces the pretrained model's WER on the target domain $4.96 \rightarrow 4.35$.
We note the data filtering techniques discussed in \cref{subsec:filtering_and_augmentation} greatly contributes to the success of self-supervision.
Without filtering, adaptation with the pseudo labels becomes much less effective, causing a WER increase on the target domain $4.35 \rightarrow 4.48$ (see \cref{tab:augmentation}).
If we also remove SpecAugment, the target domain WER further increases to $4.89$.
On the other hand, confidence filtering makes little difference since the target domain WER of the pretrained model is already low. \\[-0.1in]

\noindent \textbf{SR-RNN-T:}
To reduce on-device computation cost, the proposed SR-RNN-T loss restricts the forward/backward to a limited window of 2 frames around the most probable alignment ($b_{l} = b_{r} = 2$) according to the respective current model.
The SR-RNN-T trained model is marginally more accurate than the AR-RNN-T on both target ($4.32$) and source domain ($6.08$). \\[-0.1in]
\begin{table}[b]
  \caption{Adapting a subset of model weights. Finetuning only the key/value weights gives the best adaptation performance.}
  \label{tab:subset_params}
  \centering
  \begin{tabular}{l|rrr}
    \toprule
    \textbf{Adaptation}                     & \textbf{Params (M)}   & \textbf{Target} & \textbf{Source} \\
    \midrule                         
    All ($\circled{4}$)                     & 72.4 (100\%)          & 4.32            & 6.08           \\
    \enskip Encoder                         & 67.4 (93\%)           & 4.31            & 6.15           \\
    \quad Attention                         &  8.2 (11\%)           & 4.35            & 6.16           \\
    \enskip \quad KeyValue ($\circled{5}$)  &  4.1 (5.7\%)          & 4.32            & 5.95           \\
    \enskip Predictor                       &  1.7 (2.5\%)          & 4.68            & 6.19           \\
    \enskip Joiner                          &  3.1 (4.4\%)          & 4.70            & 6.20           \\
    \enskip Bias                            &  0.2 (0.3\%)          & 4.35            & 5.75           \\
    \bottomrule
  \end{tabular}
\end{table}

\noindent \textbf{KeyValue adaptation:}
Lastly, we experiment with finetuning only subsets of the model weights. The results are summarized in \cref{tab:subset_params}.
Finetuning only the key/value matrices of the self-attention module not only achieves a low target domain WER ($4.32$), but also prevents catastrophic forgetting on the source domain as the number of updates increases (see \cref{fig:device}).
Most importantly, key/value matrices only account for $5.7\%$ of all weights. Hence, communication cost is greatly reduced.

\subsection{Video Adaptation \label{subsec:video}}
\begin{figure}[t]
    \centering
    \includegraphics[width=1.0\linewidth]{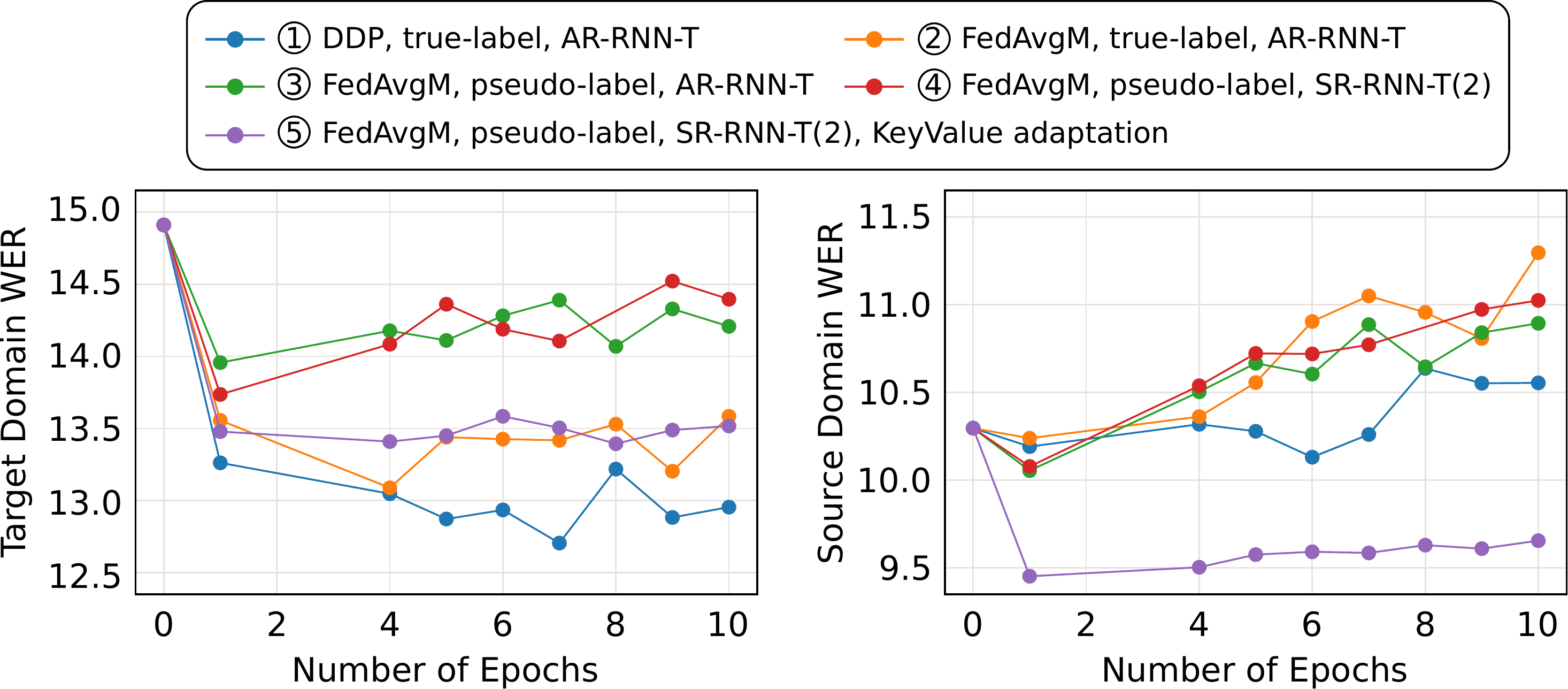}
    \caption{Convergence of Video adaptation experiments.}
    \label{fig:video}
\end{figure}
%
% TODO: We need to explain why we do video, when we are talking about on-device. We may need to do that in the abstract and/or introduction. While trying to make it not even more complicated...
%
The Video adaptation setup adapts from the source domain ``Clean'' to the target domain ``Noisy''.
The adaptation results are summarized in \cref{fig:video,tab:video}, and they confirms most of our conclusions in \cref{subsec:device}.
In particular, FL adaptation of key/value matrices with pseudo labels ($\circled{5}$) effectively reduces the WER on the target domain ($14.91 \rightarrow 13.36$).
Moreover, we have two additional learnings. \\[-0.1in]

\begin{table}[t]
  \caption{Summary of Video adaptation experiments.}
  \label{tab:video}
  \centering
  \begin{tabular}{l|rr}
    \toprule
    \textbf{Model}                                      & \textbf{Target} & \textbf{Source} \\
    \midrule
    Pretrained                                          & 14.91           & 10.30 \\
    \midrule
    $\circled{1}$ (DDP, True Labels, AR-RNN-T)          & 12.71           & 10.26 \\
    $\circled{2}$ ($\circled{1}$ + FedAvgM/BMUF)        & 13.09           & 10.36 \\
    $\circled{3}$ ($\circled{2}$ + Pseudo Labels)       & 13.89           & 10.33 \\
    $\circled{4}$ ($\circled{3}$ + SR-RNN-T)            & 13.74           & 10.08 \\
    $\circled{5}$ ($\circled{4}$ + KeyValue Adaptation) & 13.36           &  9.43 \\
    \midrule
    $\circled{6}$ ($\circled{3}$ - Filtering)           & 14.72           & 11.09 \\
    \bottomrule
  \end{tabular}
\end{table}
\noindent \textbf{Filtering:}
Since the audios in the target domain is noisy, the pseudo labels from the pretrained model have a significant error rate.
We find it necessary to apply confidence filtering using the decoding scores to remove low-confidence examples.
The optimal threshold parameter $\theta = -0.2$ removes $27\%$ of training examples.
As shown in \cref{tab:video} row \circled{6}, semi-supervised adaptation is much less effective when confidence filtering is removed, which negates most of the gains ($13.89 \rightarrow 14.72$). \\[-0.1in]

\noindent \textbf{KeyValue adaptation:}
Adapting only the key/value weights outperforms full model adaptation by large margins on both target ($13.36$ vs.\ $13.74$) and source domain ($9.43$ vs.\ $10.08$).
Moreover, the adapted model even outperforms the pretrained model on the source domain ($9.43$ vs.\ $10.30$).
This is likely because the target domain contains more diverse audio.

% \begin{table}[h]
%   \caption{Effect of confidence filtering.}
%   \label{tab:filtering}
%   \centering
%   \begin{tabular}{l|rr}
%     \toprule
%     \textbf{Model}              & \textbf{Target} & \textbf{Source} \\
%     \midrule
%     $\circled{3}$               & 13.89           & 10.33           \\
%     $\circled{3}$ - Filtering   & 14.72           & 11.09           \\
%     \bottomrule
%   \end{tabular}
% \end{table}
\section{Conclusions \label{sec:conclusion}}
This paper addressed two practical challenges for domain adaptation of an Emformer-Transducer ASR model on edge devices using Federated Learning. 
First, we show that the combination of self-restricted RNN-T loss with data augmentation and filtering techniques can improve an on-device ASR model on target domain data with full self-supervision. 
Second, we explored techniques reducing the FL compute and network communication costs by adapting only a subset of weights in the model. 
We will further explore effects of non-IID-ness and differential privacy on our proposed techniques as part of future work.

\clearpage

\bibliographystyle{IEEEtran}
\bibliography{main}

\end{document}